\documentclass{article}
\usepackage{spconf,amsmath,graphicx}
\usepackage{amssymb}
\usepackage[]{flushend}
\usepackage[hidelinks]{hyperref}
\usepackage[export]{adjustbox}

\usepackage[backend=biber,
            style=ieee,
            doi=false,
            isbn=false,
            url=false]{biblatex}
\usepackage{acronym}
\newacro{SNR}{signal-to-noise ratio}
\newacro{STFT}{short-time Fourier transform}
\newacro{STSA}{short-time spectral amplitude}
\newacro{MMSE}{minimum mean squared error}
\newacro{MSE}{mean squared error}
\newacro{TCN}{temporal convolution network}
\newacro{MAP}{\emph{maximum a posteriori}}
\newacro{A-MAP}{approximate MAP}
\newacro{ML}{maximum likelihood}
\newacro{iSTFT}{inverse STFT}
\newacro{Si-SDR}{scale-invariant signal-to-distortion ratio}
\newacro{DNS}{Deep Noise Suppression}
\newacro{ESTOI}{extended short-time objective intelligibility}
\newacro{PESQ}{perceptual evaluation of speech quality}
\newacro{DNN}{deep neural network}
\newacro{POLQA}{perceptual objective listening quality  analysis}
\newacro{MCMC}{Markov Chain Monte Carlo}
\newacro{VI}{variational inference}
\newacro{CGMM}{complex Gaussian mixture model}
\newacro{WTA}{winner-takes-all }
\newcommand{\bftab}{\fontseries{b}\selectfont}
\usepackage{booktabs}
\usepackage{multirow}
\usepackage{siunitx,etoolbox}

\usepackage{pifont}
\usepackage{amsmath}
\usepackage{mathabx}

\newcommand{\cmark}{\ding{51}}%
\newcommand{\xmark}{\ding{55}}%
\DeclareMathOperator*{\argmin}{argmin} %
\usepackage{adjustbox}
\newcommand*{\Scale}[2][4]{\scalebox{#1}{$#2$}}%
\newcommand{\rznabla}{ \resizebox{0.23cm}{0.25cm}{$\nabla$}}
\usepackage{lipsum} %
\usepackage{tikzpagenodes} %
\usepackage[moderate]{savetrees}

\title{Uncertainty Estimation in Deep Speech Enhancement\\ Using Complex Gaussian Mixture Models}

\name{Huajian Fang\thanks{We thankfully acknowledge the funding from ahoi.digital.}, Timo Gerkmann}
\address{
  Signal Processing (SP), Universität Hamburg, Germany}

\begin{document}
\ninept
\maketitle

\begin{abstract}
\vspace{-0.15cm}
Single-channel deep speech enhancement approaches often estimate a single multiplicative mask to extract clean speech without a measure of its accuracy. Instead, in this work, we propose to quantify the uncertainty associated with clean speech estimates in neural network-based speech enhancement. Predictive uncertainty is typically categorized into \emph{aleatoric uncertainty} and \emph{epistemic uncertainty}. The former accounts for the inherent uncertainty in data and the latter corresponds to the model uncertainty. Aiming for robust clean speech estimation and efficient predictive uncertainty quantification, we propose to integrate statistical complex Gaussian mixture models~(CGMMs) into a deep speech enhancement framework. More specifically, we model the dependency between input and output stochastically by means of a conditional probability density and train a neural network to map the noisy input to the full posterior distribution of clean speech, modeled as a mixture of multiple complex Gaussian components. Experimental results on different datasets show that the proposed algorithm effectively captures predictive uncertainty and that combining powerful statistical models and deep learning also delivers a superior speech enhancement performance.

\end{abstract}
\vspace{-0.15cm}

\vspace{-0.25cm}
\section{Introduction}
\label{sec:intro}
\vspace{-0.3cm}
Speech enhancement aims to recover clean speech from microphone recordings distorted by interfering noise to improve speech quality and intelligibility. The recordings are often transformed into the time-frequency domain using the \ac{STFT}, where an estimator can be applied to extract clean speech. Depending on different probabilistic assumptions, various Bayesian estimators have been presented~\cite{noisereducitonsurvey,timowienerfiltering2018}. A typical example is the Wiener filter derived based on the complex Gaussian distribution of speech and noise signals~\cite{timowienerfiltering2018}. \Acp{CGMM} have also been studied in~\cite{astudillo2013extension} to model super-Gaussian priors, which are considered a better fit for speech signals~\cite{timowienerfiltering2018}.

Today, \ac{DNN}-based approaches are the standard tool for speech enhancement, alleviating shortcomings of traditional methods. Supervised masking approaches are trained on large databases consisting of noisy-clean speech pairs and directly estimate a multiplicative mask to extract clean speech~\cite{wang2018supervised}. However, supervised \ac{DNN} approaches are typically formulated as a problem with a single output, which may result in fundamentally erroneous estimates for unseen samples, without any indication that the erroneous estimate is uncertain. This motivates us to quantify predictive uncertainty associated with clean speech estimates, which allows determining the level of confidence in the outcome in the absence of ground truth.

Predictive uncertainty is typically categorized into \emph{aleatoric uncertainty} and \emph{epistemic uncertainty}~\cite{hullermeier2021aleatoric,kendall2017uncertainties}. Aleatoric uncertainty describes the
uncertainty of an estimate due to the intrinsic randomness of noisy observations. For speech enhancement, it originates from the stochastic nature of both speech and noise. Epistemic uncertainty (also known as \emph{model uncertainty}) corresponds to the uncertainty of the \ac{DNN}
parameters~\cite{kendall2017uncertainties}. H{\"u}llermeier et al.~\cite{hullermeier2021aleatoric} provide a general introduction to uncertainty modeling. To quantify aleatoric uncertainty, the dependency between input and output can be modeled stochastically using a speech posterior distribution and enable the \ac{DNN} to estimate the statistical moments of this distribution. While the predictive mean is a target estimate, the associated variance can be used to measure aleatoric uncertainty~\cite{kendall2017uncertainties}. Previous work has implicitly or explicitly explored the uncertainty of aleatoric nature in the context of \ac{DNN}-based speech enhancement. Chai et al.~\cite{chai2019generalizedgaussian} have proposed a generalized Gaussian distribution to model prediction errors in the log-spectrum domain. Siniscalchi~\cite{siniscalchi2021vector} has proposed to use a histogram distribution to approximate the conditional speech distribution, but with a fixed variance assumption, thus failing to capture input-dependent uncertainty. Our previous work~\cite{fang2022integrating} allows capturing aleatoric uncertainty based on the complex Gaussian posterior. In contrast, quantifying epistemic uncertainty in the context of \ac{DNN}-based speech enhancement approaches to account for model's imperfections remains relatively unexplored. In computer vision and deep learning, epistemic uncertainty is usually captured using approximate Bayesian inference. For instance, variational inference can approximate the exact posterior distribution of \ac{DNN} weights with a tractable distribution~\cite{kendall2017uncertainties,blundell2015weight}. At testing time, multiple sets of \ac{DNN} weights can be obtained by sampling from an approximate posterior network weight distribution, thus producing multiple different output predictions for each input sample. Epistemic uncertainty captures the extent to which these weights vary given input data, which can be empirically quantified by the variance in these output predictions~\cite{kendall2017uncertainties}. However, its computational effort is proportional to the number of sampling passes. This renders those approaches impractical for devices with limited computational resources or strict real-time constraints.

In this work, we propose to integrate statistical \acp{CGMM} into a deep speech enhancement framework, so that we can combine the powerful nonlinear modeling capabilities provided by neural networks with super-Gaussian priors as a way to improve the robustness of the algorithm as well as to capture predictive uncertainty. More specifically, we propose to train a \ac{DNN} to estimate the full posterior distribution of clean speech, modeled as a mixture of multiple complex Gaussian components. The one-to-many mapping based on the~\ac{CGMM} enables the \ac{DNN} to make multiple reasonable hypotheses,
thus increasing the robustness against adverse acoustic scenarios. At the same time, in addition to clean speech estimates, the proposed framework featuring one-to-many mappings allows capturing both aleatoric uncertainty and epistemic uncertainty without extra computational costs. 
Furthermore, we propose a pre-training scheme to mitigate the mode collapse problem often observed in mixture models, resulting in improved clean speech estimation. Finally, we adapt and employ a gradient modification scheme to effectively stabilize the training of our mixture model.%

Note that previous work by Kinoshita et al.~\cite{kinoshita2017deep} also seeks to output multiple hypotheses to avoid deterministic mappings. Our work is different in two main aspects. First, Kinoshita et al. model the logarithm Mel-filterbank features using real Gaussian mixture models, while we follow the prior \ac{CGMM} of speech and noise spectral coefficients. Second, the \ac{DNN} outputs in~\cite{kinoshita2017deep} serve as the basis for an additional statistical model-based enhancement method, while we target to obtain clean speech estimates directly via \acp{DNN} in an end-to-end fashion.

\vspace{-0.2cm}
\section{Signal Model}
\label{sec:signlmodel}
\vspace{-0.25cm}
We consider a single-channel speech enhancement problem in which clean speech is distorted by additive noise. In the \ac{STFT} domain, the noisy signal is given by
\vspace{-0.1cm}
\begin{equation}
   X_{ft} = S_{ft} + N_{ft},
\end{equation}
where $S_{ft}$ and $N_{ft}$ denote the speech and noise complex coefficients at the frequency bin~$f\in \{1,\dotsc,F\}$ and the time frame~$t\in \{1,\dotsc,T\}$. We model the speech and noise signals as mixtures of zero-mean complex Gaussian distributions~\cite{astudillo2013extension}:
\begin{equation}
\vspace{-0.1cm}
 \label{eq:prior}
    S_{ft} \sim \sum_{i=1}^I \;\Omega(i)\;\mathcal{N}_\mathbb{C}(0,\;\sigma^{2}_{i,ft}), \hspace{0.3cm}
    N_{ft} \sim \sum_{j=1}^J\; \Omega(j)\;\mathcal{N}_\mathbb{C}(0,\;\sigma^{2}_{j,ft}) \, .
\vspace{-0.1cm}
\end{equation}
The speech mixture weights $\Omega(i)$ sum to one, and the same applies to the noise mixture weights $\Omega(j)$. 
The likelihood $p(X_{ft}|S_{ft})$ follows a complex Gaussian mixture distribution centered at $S_{ft}$, given by 
\begin{equation}
\vspace{-0.1cm}
    \label{eq:likelihood}
    p(X_{ft}|S_{ft}) = \sum_{j=1}^J \; \Omega(j)\;\frac{1}{\pi \sigma^2_{j,ft}} \exp\left(-\frac{|X_{ft}-S_{ft}|^2}{\sigma^2_{j,ft}}\right) \, .
\vspace{-0.1cm}
\end{equation}
Given the speech prior in~(\ref{eq:prior}) and the likelihood distribution in~(\ref{eq:likelihood}), one can apply Bayes' theorem to determine the posterior distribution of speech as follows~\cite{astudillo2013extension}
\begin{equation}
\vspace{-0.1cm}
     \label{eqn:posteriorcoplex}
    p(S_{ft}|X_{ft}) =\sum_{i=1}^I\sum_{j=1}^J\; \Omega(i,j|X_{ft})\frac{1}{\pi\lambda_{ij,ft}} \exp{\left(-\frac{|S_{ft} - W^{\text{WF}}_{ij,ft}X_{ft}|^2}{\lambda_{ij,ft}}\right)} \, ,
\vspace{-0.1cm}
\end{equation}
where $W^{\text{WF}}_{ij,ft}=\frac{\sigma_{i,ft}^2}{\sigma_{i,ft}^2 + \sigma_{j,ft}^2}$ and $\lambda_{ij,ft} = \frac{\sigma_{i,ft}^2\sigma_{j,ft}^2}{\sigma_{i,ft}^2 + \sigma_{j,ft}^2}$ are the Wiener filter and the posterior's variance of the mixture Gaussian pair $(i,j)$, respectively. $\Omega(i,j|X_{ft})$ denotes the posterior's mixture weights with the same sum-to-one constraint. The variance $\lambda_{ij,ft}$ for the mixture pair $(i,j)$ can be interpreted as a measure of uncertainty for the Wiener estimate $\widetilde{S}_{ij,ft} = W^{\text{WF}}_{ij,ft} X_{ft}$~\cite{timowienerfiltering2018}. 

Given an input noisy signal, multiple complex Gaussian components can be combined by computing the expectation of the posterior \ac{CGMM}, yielding the clean speech estimate
\begin{equation}
\vspace{-0.1cm}
    \begin{split}
    \mathbb{E}(S_{ft}|X_{ft}) =  \int S_{ft} \; p(S_{ft}|X_{ft}) dS_{ft}=\sum_{i=1}^{I}\sum_{j=1}^{J}\Omega(i,j|X)\widetilde{S}_{ij,ft}\ .
    \end{split}
    \vspace{-0.1cm}
\end{equation}

The mixture density model possesses the advantage of being able to approximate an arbitrary density function with a sufficient number of components~\cite{goodfellow2016deep}, which provides a good fit for modeling, e.g., super-Gaussian characteristics of the speech coefficients. In this work, we propose to embed the \ac{CGMM} into a \ac{DNN} framework in order to additionally take advantage of their non-linear modeling capacities, as will be shown next. 

\vspace{-0.15cm}
\section{Joint estimation of clean speech\\ and predictive uncertainty}
\label{sec:jointestimation}
\vspace{-0.25cm}
Instead of relying on traditional power spectral density tracking algorithms, we can leverage neural networks to directly estimate a mixture of Wiener filters to recover clean speech. Furthermore, it is also possible to optimize the neural network based on the speech posterior distribution~(\ref{eqn:posteriorcoplex}), so that not only the Wiener filter but also the variance of each mixture pair can be estimated. By taking the negative logarithm  and averaging over the time-frequency bins, we can obtain the following optimization problem
\vspace{-0.1cm}
\begin{equation}
    \begin{split}
  \widetilde{W}^{\text{WF}}_{l,ft},&\;\widetilde{\lambda}_{l,ft},\;\widetilde{\Omega}_{l,ft} =\argmin_{W^{\text{WF}}_{l,ft},\lambda_{l,ft}, \Omega_{l,ft}} 
    \overbrace{-\frac{1}{FT}\sum_{f,t} \log\left(
    \sum_{l=1}^{L}\;\exp\left(
   \Theta_{l,ft}
    \right)
    \right) 
    }^{\mathcal{L}_{p(S|X)}^{\text{CGMM}}} \ , \\
   &\Theta_{l,ft} = 
    \log\left(\Omega(l|X_{ft})\right)-\log(\lambda_{l,ft})-\frac{|S_{ft} - W^{\text{WF}}_{l,ft}X_{ft}|^2}{\lambda_{l,ft}} \ ,  
    \end{split}
    \label{eqn:logposterior}
    \vspace{-0.3cm}
\end{equation}
where $l\in \{1,\dotsc,L\}$ indexes a mixture pair $(i,j)$ in~(\ref{eqn:posteriorcoplex}), i.e., $L=I\times J$. $\widetilde{W}^{\text{WF}}_{l,ft}$ and $\widetilde{\lambda}_{l,ft}$ denote estimates of the Wiener filter and its associated uncertainty. The \ac{CGMM} and the corresponding loss function can be viewed as a generalization of the uni-modal Gaussian assumption, which in turn is a generalization of the \ac{MSE} loss function. In the limiting case $L=1$ (i.e., $I=J=1$), $\mathcal{L}_{p(S|X)}^{\text{CGMM}}$ degenerates into the generic complex Gaussian with a single mean $W^{\text{WF}}_{ft}$ and variance $\lambda_{ft} $, such that
\vspace{-0.15cm}
\begin{equation}
    \begin{split}
    \widetilde{W}^{\text{WF}}_{ft}, \widetilde{\lambda}_{ft} = 
    \argmin_{W^{\text{WF}}_{ft},\lambda_{ft}} \underbrace{\frac{1}{FT}\sum_{f,t} \log(\lambda_{ft})  + \frac{|S_{ft} - W^{\text{WF}}_{ft} X_{ft}|^2}{\lambda_{ft}}}_{\mathcal{L}_{p(S|X)}^{\text{CG}}} \ .
    \end{split}
    \label{eqn:logposterior1}
\vspace{-0.2cm}
\end{equation}
Furthermore, by assuming a constant uncertainty for all time-frequency bins and refraining from optimizing for it,  $\mathcal{L}_{p(S|X)}^{\text{CG}}$ degenerates into the commonly-used \ac{MSE} loss~\cite{braun2021consolidated}:
\vspace{-0.15cm}
\begin{equation}
\mathcal{L}_{\text{MSE}} = \frac{1}{FT}\sum_{f,t}|S_{ft}-W^{\text{WF}}_{ft}X_{ft}|^2 \, .
\label{eqn:mse}
\vspace{-0.2cm}
\end{equation}

\begin{figure}[t]
  \centerline{\includegraphics[width=9.5cm,height=3.9cm,left]{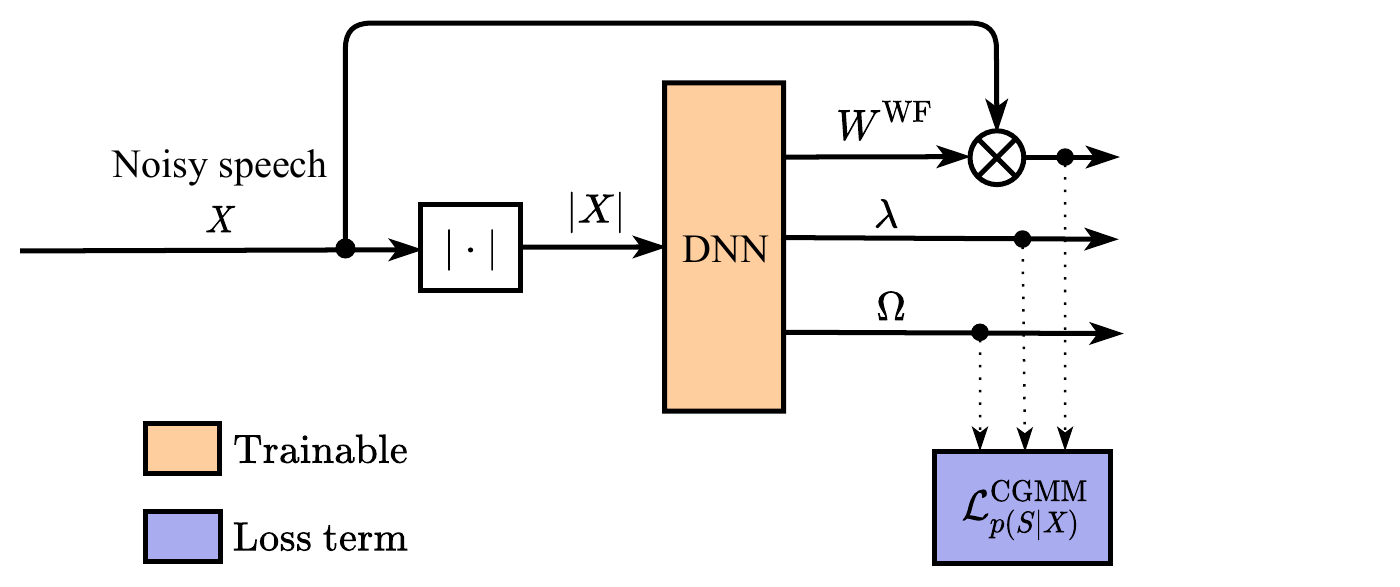}}
  \vspace{-0.3cm}
\caption{Block diagram of the \ac{DNN}-based predictive uncertainty estimation.}
\vspace{-0.3cm}
\label{fig:cgmmdiagram} 
\end{figure}

In this work, we depart from the uni-modal Gaussian and the constant uncertainty assumption. Alternatively, we propose to map the input to multiple hypotheses by training a DNN with the negative log-posterior~$\mathcal{L}_{p(S|X)}^{\text{CGMM}}$, so that we can leverage the better modeling capabilities of the multi-modal distribution and also enable the \ac{DNN} to quantify the overall predictive uncertainty. 
Furthermore, incorporating the variances associated with the Wiener estimates enables the adjustment of the weighting of the residual loss, as interpreted in~\cite{kendall2017uncertainties}, improving the robustness of the network to adverse inputs. As all time-frequency bins are treated independently in~(\ref{eqn:posteriorcoplex}), the indices $ft$ will be omitted hereafter wherever possible. 

We can compute the posterior's variance~\cite[Section 5]{bishop2006pattern}, $\text{Var}(S|X)$, to quantify the overall (squared) uncertainty in clean speech estimates originating from different aspects. With the law of total variance, the posterior's variance can be decomposed into~\cite{choi2018uncertainty}:
\begin{equation}
\begin{split}
 \resizebox{0.89\hsize}{!}{$\text{Var}(S|X) =  \underbrace{\sum_l^L\Omega(l|X)\ \lambda_l}_{\displaystyle  \mathbb{E}_{l\sim \Omega(l|X)}[\text{Var}(S|X,\;l)]} + \ \ \underbrace{\sum_l^L\;\Omega(l|X)\;\left|W^{\text{WF}}_{l} X - \mathbb{E}(S|X) \right|^2}_{\displaystyle \small\text{Var}_{l\sim \Omega(l|X)}(\mathbb{E}[S|X,\;l])}$}.
\end{split}
\vspace{-0.1cm}
\end{equation} 
The inherent uncertainty associated with the $l$-th Gaussian component in the outcome is given by $\text{Var}(S|X,\;l) = \lambda_l$ and aleatoric uncertainty is then quantified as the expectation of variance components $\mathbb{E}_{l\sim \Omega(l|X)}[\text{Var}(S|X,\;l)]$ following the interpretation in~\cite{kendall2017uncertainties,choi2018uncertainty}. Epistemic uncertainty can be captured using multiple output predictions, which can be achieved here by the mixture of Wiener estimates, thus circumventing the need for an expensive sampling process. For this, one can compute the variance of the conditional expectation, resulting in the epistemic uncertainty estimate~$\text{Var}_{l\sim \Omega(l|X)}(\mathbb{E}[S|X,\;l])$. Fig~\ref{fig:cgmmdiagram} depicts an overview of this approach.

Probability density estimation is a non-trivial task. Our preliminary experiments using $\mathcal{L}_{p(S|X)}^{\text{CGMM}}$ directly as the loss function have shown numerical instabilities during training. To overcome this, a gradient modification scheme inspired by~\cite{seitzer2022pitfalls} is adapted and employed. Furthermore, \acp{DNN} optimized based on $\mathcal{L}_{p(S|X)}^{\text{CGMM}}$ are not guaranteed to exploit the multi-modality of the mixture model, i.e., the multiple hypotheses may converge to the same estimate (collapse to a single mode). We propose to handle this using a pre-training technique based on the \ac{WTA} scheme~\cite{guzman2012multiple}.

\vspace{-0.1cm}
\subsection{Gradient modification scheme}
\label{ssec:dependence}
\vspace{-0.12cm}
The optimization of \acp{DNN} with a uni-modal Gaussian~(e.g.,~(\ref{eqn:logposterior1})) as the loss function using stochastic gradient descent shows a high dependence of the gradient on the variance, which is known to cause optimization instabilities~\cite{takahashi2018student,seitzer2022pitfalls}. This can be particularly problematic in our \ac{CGMM} involving multiple complex Gaussian components. It can be seen by computing the gradients of the exponential term $\Theta_l$ in~(\ref{eqn:logposterior}) with respect to the $l$-th Wiener filter and associated variance, shown as follows  
\begin{equation}
\vspace{-0.1cm}
 \resizebox{0.89\hsize}{!}{$ \rznabla_{\Scale[0.65]{W_l^{\text{WF}}}}\Theta_{l} = \frac{2\text{Re}\{-S\widebar{X}+W_l^{\text{WF}}|X|^2\}}{\lambda_l}\;,\;
 \rznabla_{\lambda_l}\Theta_{l} = \frac{\lambda_l-|S-W_l^{\text{WF}}X|^2}{\lambda_l^2}
 $},
\label{eq:gradient_wf}
\vspace{-0.1cm}
\end{equation}
where the $\text{Re}\{\cdot\}$ operation returns the real part and $\widebar{\cdot}$ denotes the complex conjugate. A recent analysis of the real-valued Gaussian assumption by Seitzer et al.~\cite{seitzer2022pitfalls} showed that the dependence of the gradient on the variance can be reduced by modifying the gradient based on the variance value. Inspired by this, here we extend it to the mixture model, which can be achieved by introducing a weighting term~$\lambda^{\beta_l}$ to each complex Gaussian component in the loss~(\ref{eqn:logposterior}):
\vspace{-0.1cm}
\begin{equation}
   \widetilde{W}^{\text{WF}}_{l,ft},\;\widetilde{\lambda}_{l,ft}=     \argmin_{W^{\text{WF}}_{l,ft},\lambda_{l,ft}} -\frac{1}{FT}\sum_{f,t} 
    \log\left(
    \sum_{l=1}^{L}\;
    \exp (\text{sg}[\lambda^{\beta_l}]\,\Theta_{l,ft})
    \right) 
    \, ,  
\label{eqn:logposterior_beta}
\vspace{-0.1cm}
\end{equation}
where $\text{sg}[\cdot]$ denotes the stop gradient operation, which allows $\lambda^{\beta_l}$ to act as an input-dependent adaptive factor on the gradient. The parameter $\beta_l\in [0,1]$ controls how much the gradient depends on the $l$-th variance. As a result, the gradients are modified to
\begin{equation}
\vspace{-0.1cm}
 \resizebox{0.89\hsize}{!}{$
  \rznabla^\prime_{\Scale[0.65]{W_l^{\text{WF}}}}\Theta_{l} = \frac{2\text{Re}\{-S\widebar{X}+W_l^{\text{WF}}|X|^2\}}{\lambda_l^{1-\beta_l}}\;,\;
  \rznabla^\prime_{\lambda_l}\Theta_{l} = \frac{\lambda_l-|S-W_l^{\text{WF}}X|^2}{\lambda_l^{2-\beta_l}}
  $}\;.
\label{eq:gradient_wf_beta}
\vspace{-0.1cm}
\end{equation}
Experimentally, we find that the modification in~(\ref{eqn:logposterior_beta}) is effective in addressing instability problems during the training of the probabilistic mixture models.

\vspace{-0.15cm}
\subsection{WTA pre-training scheme}
\label{sec:wta}
\vspace{-0.1cm}

In order to obtain diverse predictions, we propose a pre-training scheme based on the \ac{WTA} loss~\cite{guzman2012multiple} to introduce a competition mechanism among the output layers. The concept was originally presented by Guzman-Rivera et al.~\cite{guzman2012multiple} for support vector machines to produce multiple outputs, and later generalized to the context of \acp{DNN}~\cite{lee2016stochastic,ilg2018uncertainty, makansi2019overcoming,panousis2019nonparametric}. We apply the pre-training procedure to a \ac{DNN} which outputs multiple masks to generate clean speech estimates, i.e., it is equivalent to the \ac{CGMM} consisting of only the mixture of Wiener estimates. To prompt a network to output diverse hypotheses based on a single ground-truth, the gradient is backpropagated through the top $K$ of the $L$ output predictions at each iteration~\cite{makansi2019overcoming}:
\vspace{-0.1cm}
\begin{equation}
\begin{split}
   \mathcal{L}_{\text{WTA}} = \frac{1}{K}\sum_{k=1}^K\mathcal{L}_{\text{MSE}}(W_k^{\text{WF}}X,\;\; S),
\end{split}
\vspace{-0.2cm}
\label{eq:wta}
\end{equation}
where the top-$K$ winners are selected based on the \ac{MSE} measure, indexed by $k$. Following~\cite{makansi2019overcoming}, we start with $K=L$, and gradually halve the number of selections until reaching $K=1$. The competition mechanism prompts the \ac{DNN} to output diverse clean speech estimates to capture the model's uncertainty, which is expected to alleviate the mode collapse problem to some extent. Previous work has proposed feeding these predictions into a post-processing network to perform distribution fitting~\cite{ilg2018uncertainty, makansi2019overcoming}, while here we propose to use it to initialize the \ac{CGMM} (except for the output layers that estimates the mixing coefficient and variance of each Gaussian component) to strengthen clean speech estimation without introducing any additional parameters.

\vspace{-0.15cm}
\section{Experiments}
\label{sec:experiment}
\vspace{-0.15cm}
\subsection{Dataset}
\vspace{-0.15cm}
We randomly select 80 and 20 hours from the \ac{DNS} Challenge dataset~\cite{reddy2020interspeech} for training and validation, respectively. The \ac{SNR} is uniformly sampled between -5 dB and 20 dB. We evaluate the model on two unseen datasets. The first is the non-reverb synthetic test set released by DNS Challenge, which is created by mixing speech signals from~\cite{pirker2011pitch} with noise from 12 categories~\cite{reddy2020interspeech}, at \acp{SNR} ranging from 0~dB and 25~dB. We synthesize the second test set by mixing speech samples from WSJ0~(\texttt{si\_et\_05})~\cite{garofolo1993csr} and noise samples from CHiME~(\texttt{cafe}, \texttt{street}, \texttt{pedestrian}, and \texttt{bus})~\cite{chime3dataset} at \acp{SNR} randomly chosen from \{-10~dB, -5~dB, 0~dB, 5~dB, 10~dB\}.
\vspace{-0.2cm}

\begin{table}[t]
\centering
\resizebox{1.02\columnwidth}{!}{
\begin{tabular}{|c||c|c||c|c|c|c|}
\hline
\multirow{2}{*}{Methods}&
\multirow{2}{*}{Ale.}& \multirow{2}{*}{Epi.}& \multicolumn{3}{c|}{SNR} & \multirow{2}{*}{Average}\\  \cline{4-6}
& & &  \textless 6 dB & 6-12 dB & \textgreater 12 dB & \\
\hline
Noisy & - & - & 1.33/0.74 & 1.52/0.81 & 2.00/0.91 & 1.58/0.81\\ \hline
Baseline WF & \xmark & \xmark & 2.05/0.85 & 2.53/0.91 & 3.03/0.95 & 2.48\textbf{}/0.90\\ \hline
Prop. CGMM1 & \cmark & \xmark & 2.20/\bftab0.86 & 2.66/\bftab0.92 & 3.13/\bftab0.96 & 2.61/\bftab0.91 \\ \hline
Prop. CGMM4-cons & \xmark & \cmark & 2.16/\bftab0.86 & 2.65/\bftab0.92 & 3.17/\bftab0.96 & 2.60/\bftab0.91 \\ \hline
Prop. CGMM4 &\cmark &\cmark & 2.21/\bftab0.86 & 2.68/\bftab0.92 & 3.13/\bftab0.96 & 2.62/\bftab0.91\\ \hline
Prop. CGMM4-pre & \cmark & \cmark& \bftab2.22/0.86 & \bftab2.78/0.92 & \bftab3.24/\bftab0.96 & \bftab 2.69/\bftab0.91\\ \hline
\end{tabular}
}\vspace{-0.2cm}
  \caption{Average performance on DNS non-reverb test set. The values are given in PESQ/ESTOI. \emph{Ale.}: Aleatoric; \emph{Epi.}: Epistemic;  \emph{Prop.}: Proposed.}
  \label{dnsresults}
  \vspace{-0.45cm}
\end{table}

\vspace{-0.1cm}
\subsection{Experimental settings}
\vspace{-0.15cm}
We compute the \ac{STFT} using a 32~ms Hann window and 50\% overlap, at a sampling rate of 16~kHz. For a fair comparison, we base all experiments on a plain U-Net architecture adapted from~\cite{Jansson2017SingingVS, tan2018convolutional}. The architecture has skip connections between the encoder and the decoder and consists of multiple identical blocks, of which each consists of: 2D convolution layer + instance normalization + Leaky ReLU with slope 0.2. The model processes the inputs of dimension $(T, F)$, with the kernel size $(5, 5)$, stride $(1, 2)$, and padding $(2, 2)$. The encoder is comprised of 6 blocks that increase the feature channel from $1$ to $512$ progressively~($1-16-32-64-128-256-512$), and then the decoder reduces it back to 16~($512-256-128-64-32-16-16$), followed by a $1\times1$ convolution layer that outputs a single mask of the same shape as the input when performing point estimation or outputs $L$ pairs of masks, variance estimates, and mixture weights when applying the \ac{CGMM}. We set $I$ and $J$ to 2 in~(\ref{eqn:posteriorcoplex}) , resulting in $L=4$. We set $\beta_l$ to 0.5 for $l\in\{1,\cdots, L\}$ following~\cite{seitzer2022pitfalls}. 

The models are trained using the Adam optimizer with a learning rate of $10^{-3}$, which is halved if the validation loss does not decrease for consecutive 3 epochs. Early stopping with a patience of 10 epochs is used.  The batch size is 64; the weight decay factor is set to $0.0005$. The CGMM can be optionally pre-trained based on the \ac{WTA} loss as described in~Section~\ref{sec:wta}. Since it is not straightforward to determine a validation loss for the \ac{WTA} mechanism, we train the model for 125 epochs with the initial learning rate $10^{-3}$, and then halve it every 5 epochs when it is greater than $10^{-6}$. We halve the number of winners after every 25 epochs, from $K=4$ to $K=2$, eventually reaching $K=1$, while $K$ remains at $1$ for the rest of the training process. The \ac{CGMM} is then fine-tuned with an initial learning rate of $10^{-5}$ and the same decay and stopping schemes. Note that the proposed gradient modification scheme described in~Section~\ref{ssec:dependence} is employed to stabilize the training of all \ac{CGMM}-based networks.

Finally, the following deep algorithms are evaluated:
\begin{enumerate}
    \item \emph{Baseline WF} refers to a single Wiener filter trained with the loss~(\ref{eqn:mse}). 
    \item \emph{CGMM1} refers to the \ac{CGMM} with $L=1$ (i.e., $I=J=1$) trained using the loss~(\ref{eqn:logposterior_beta}). It outputs a single Wiener filter and variance, thus modeling only aleatoric uncertainty.
    \item \emph{CGMM4} denotes the \ac{CGMM} with $L=4$ trained using the loss~(\ref{eqn:logposterior_beta}), which captures both aleatoric and epistemic uncertainties.
    \item \emph{CGMM4-\textbf{cons}} assumes a \textbf{cons}tant variance for \ac{CGMM}4 and refrains from optimizing for it~($\lambda_{l,ft}=1$), capturing epistemic uncertainty through a mixture of Wiener estimates.
    \item \emph{CGMM4-\textbf{pre}} refers to the \ac{CGMM}4 \textbf{pre}-trained with the \ac{WTA} loss.
    \vspace{-0.15cm}
\end{enumerate}

\vspace{-0.05cm}
\subsection{Metrics}
\vspace{-0.15cm}
We present speech enhancement results in terms of \ac{PESQ} and \ac{ESTOI}. We use a sparsification plot~\cite{wannenwetsch2017probflow,ilg2018uncertainty} to quantitatively evaluate the captured uncertainty. The sparsification plot illustrates the correlation between the uncertainty measures and the true errors. As a first step, the errors of the spectral coefficients are ranked according to their corresponding uncertainty measures. For well-calibrated uncertainty estimates, when the time-frequency bins with large uncertainties are removed the residual error should decrease. Accordingly, the root mean squared error~(RMSE) can be plotted versus the fraction of the time-frequency bins removed. Ranking the true errors by their own values yields a lower bound for the sparsification plot, referred to as the \emph{oracle curve}. When the uncertainty estimates and the errors are perfectly correlated, the sparsification plot and the oracle curve overlap.

\begin{figure}[t]
  \centerline{\includegraphics[width=9.2cm,height=5cm]{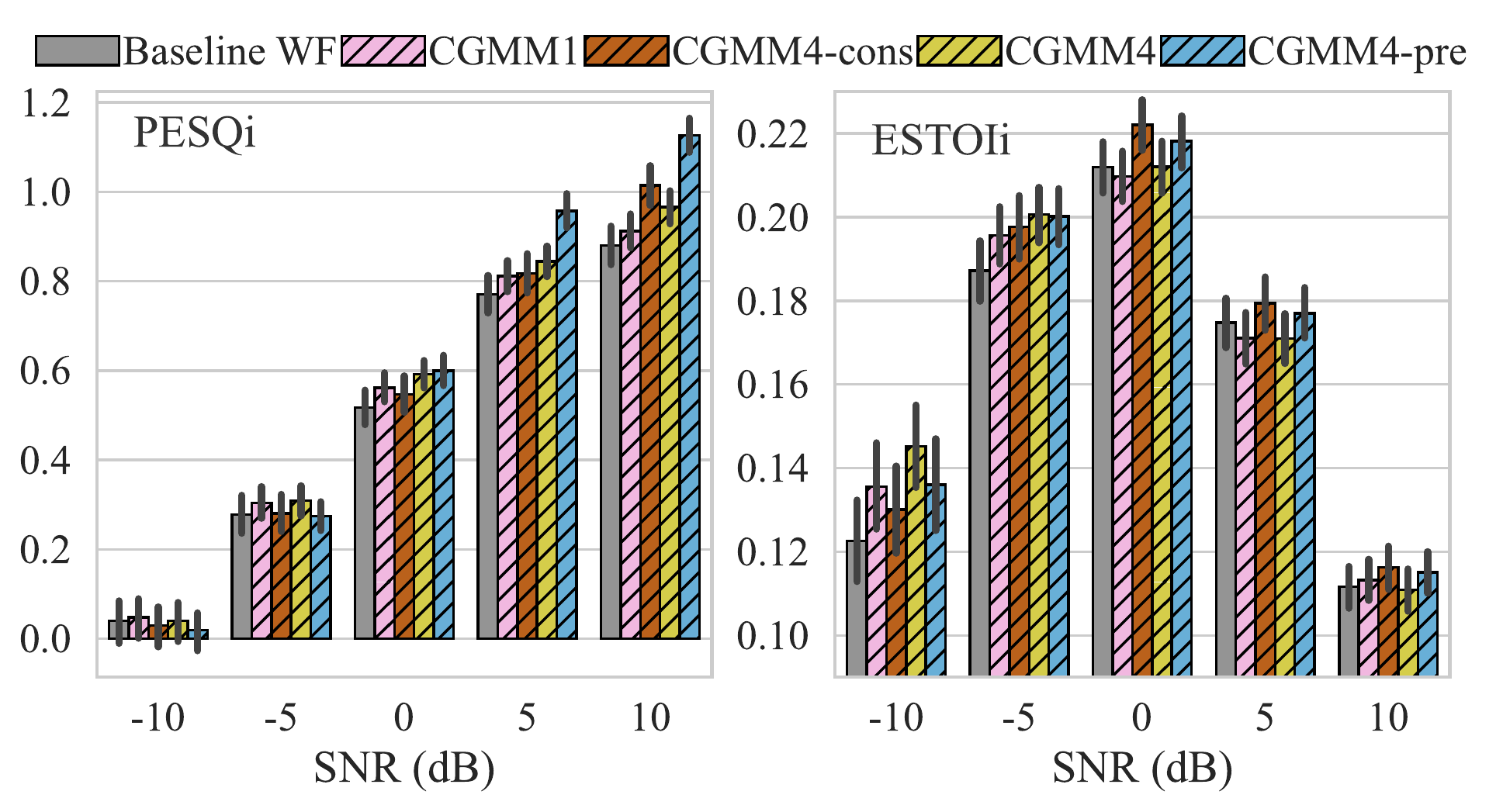}}
  \vspace{-0.4cm}
\caption{Performance improvement obtained on the  WSJ0-CHiME test set. PESQi denotes PESQ improvement relative to noisy mixtures. The same definition applies to ESTOIi. The marker denotes the mean value and the vertical bar indicates the 95\%-confidence interval.}
\vspace{-0.5cm}
\label{fig:wsjchime} 
\end{figure}

\vspace{-0.15cm}
\subsection{Results}
\vspace{-0.12cm}
In Table~\ref{dnsresults}, we present average evaluation results on the DNS non-reverb test set. It can be observed that the proposed framework considering either aleatoric uncertainty~(CGMM1) or epistemic uncertainty~(CGMM4-cons) outperforms the point estimation baseline, demonstrating the advantages of modeling uncertainty associated with the clean speech estimates in speech enhancement. Comparing CGMM4 with CGMM1 and CGMM4-cons, the benefits of the modeling both aleatoric and epistemic uncertainties using the multi-modal posterior distribution is not evident. This may be attributed to the fact that training a model based on~(\ref{eqn:logposterior_beta}) is not guaranteed to explore the multi-modal modeling capacities of the mixture model. However, this can be largely mitigated by the proposed pre-training scheme, as indicated by the higher \ac{PESQ} scores of CGMM4-pre. 

Fig.~\ref{fig:wsjchime} shows the improvements of \ac{PESQ} and \ac{ESTOI} relative to the noisy mixtures of the synthetic WSJ0-CHiME test set. We observe larger \ac{PESQ} improvements for the mixture models especially at high input \acp{SNR}. In particular, CGMM4-pre yields the highest \ac{PESQ} improvements, indicating that promoting diverse predictions in the mixture model improves generalization capacities for speech enhancement. Furthermore, it can be observed that the models accounting for uncertainty lead to larger \ac{ESTOI} improvements at low input \acp{SNR}, which again demonstrates the benefits of integrating statistical models into deep speech enhancement as well as modeling uncertainty.

\begin{figure}[t]
  \centerline{\includegraphics[width=8.5cm,height=5.55cm]{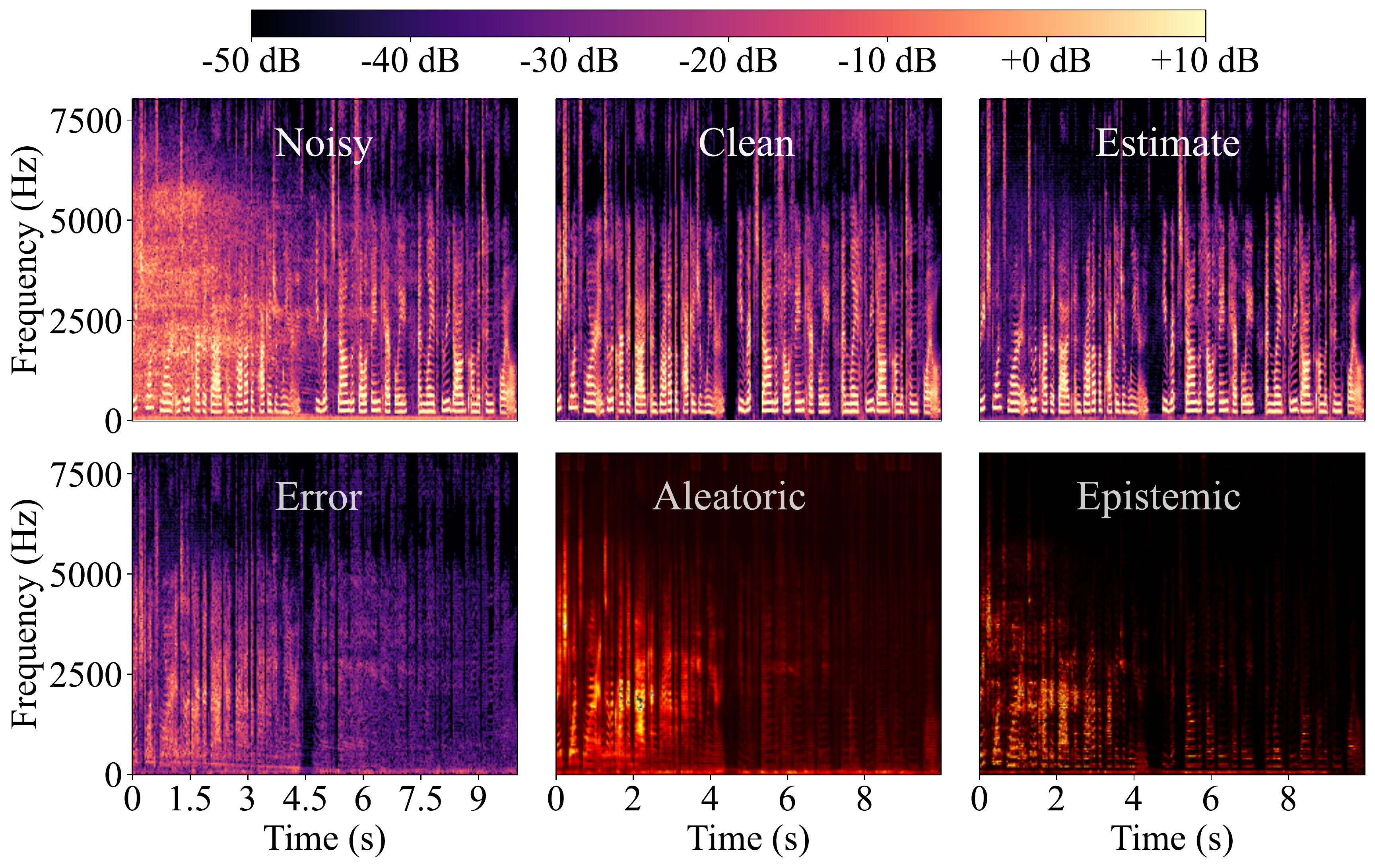}}
  \vspace{-0.2cm}
\caption{The uncertainty is visualized as a heatmap, where black indicates low uncertainty and brighter colors indicate higher uncertainty.}%
\vspace{-0.2cm}
\label{fig:spectrogram} 
\end{figure}

\begin{figure}[t]\centerline{\includegraphics[width=5cm,height=3.6cm]{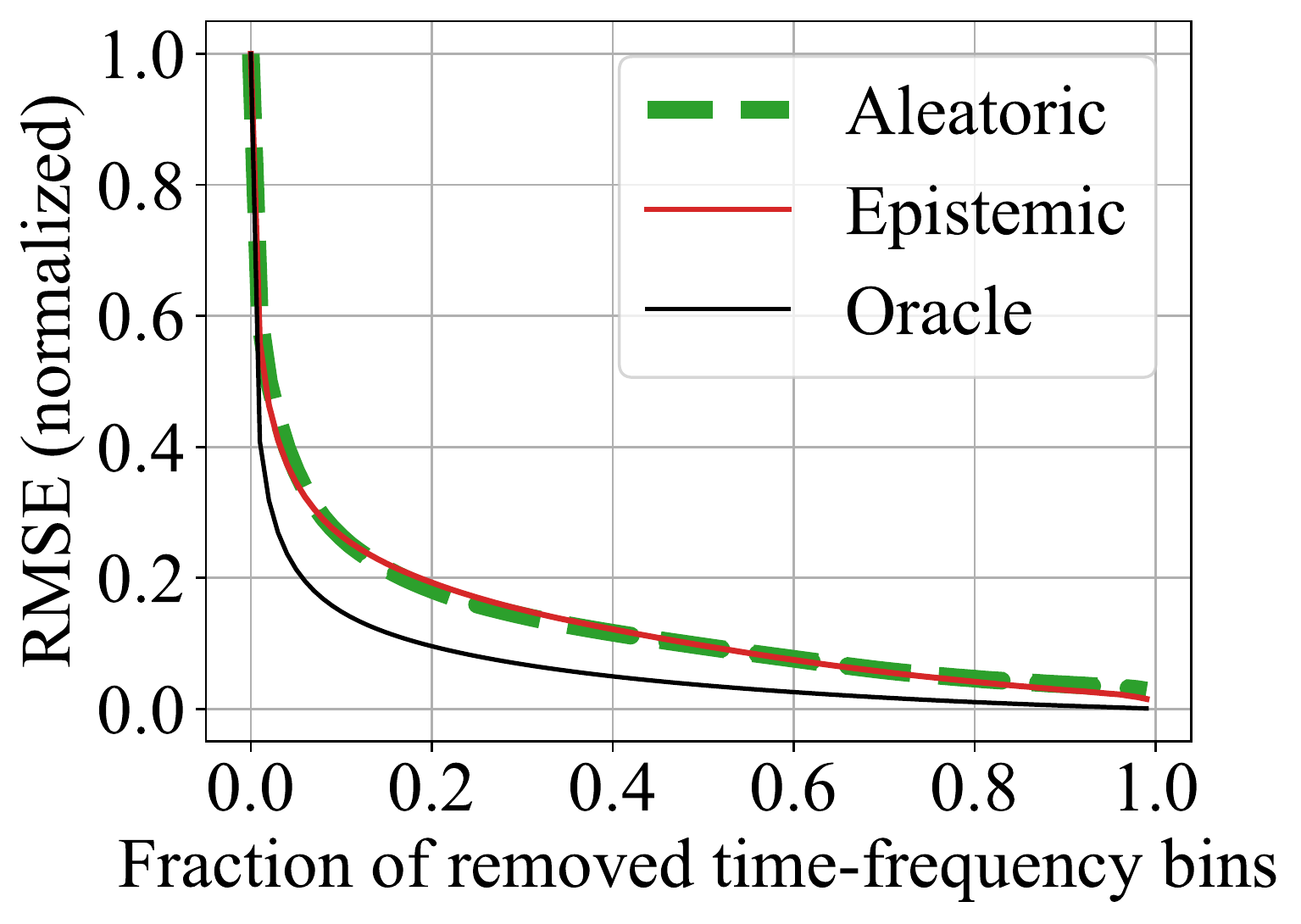}}
  \vspace{-0.2cm}
\caption{Sparsification plots created based on all audio samples of the DNS reverb-free synthetic test set. The smaller distance to the oracle curve indicates a more accurate uncertainty estimation.}
\vspace{-0.3cm}
\label{fig:sparsification} 
\end{figure}
In addition to improving performance, enabling the \acp{DNN} to quantify predictive uncertainty is essential to determine how informative a clean speech estimate is without knowing ground-truth (which we do not have access to in practice). Therefore, we evaluate the captured predictive uncertainty in CGMM-pre qualitatively and quantitatively. Fig.~\ref{fig:spectrogram} shows the spectrograms of an example utterance from the DNS test set. By computing the absolute difference between the clean reference and estimated spectral coefficients, we can measure the prediction error (visualized in the first figure of the second row). It can be observed that both types of uncertainty are closely related to the estimation error, i.e., the model outputs large uncertainties when large errors are produced (e.g., the first 4 seconds of the example utterance). This association is further reflected in Fig.~\ref{fig:sparsification}, where we observe that both sparsification plots are monotonically decreasing and are close to the oracle curve, implying that both types of uncertainties accurately reflect regions where speech prediction is difficult.

\vspace{-0.15cm}
\section{Conclusion}
\label{sec:conclusion}
\vspace{-0.1cm}

In this paper, we have proposed a deep speech enhancement framework to jointly estimate clean speech and quantify predictive uncertainty, based on the statistical \ac{CGMM}. By estimating the parameters of the full speech posterior distribution involving multiple complex Gaussian components, we can effectively capture both aleatoric and epistemic uncertainties with a single forward pass, circumventing the need for expensive sampling. In addition, the potential of the mixture models can be better exploited if we promote diverse predictions and mitigate the mode collapse problem using the proposed pre-training scheme. Eventually, evaluation results in terms of instrumental measures have demonstrated the considerable advantages of combining powerful statistical models and deep learning compared to directly predicting a point estimate. Our reliable uncertainty estimates can enable interesting future work. For instance, the uncertainty of aleatoric nature can guide multi-modality fusion~\cite{tellamekala2022cold}, while epistemic uncertainty capturing the model's ignorance~\cite{hullermeier2021aleatoric} can be used to design a uncertainty-driven training mechanism to improve, e.g., domain adaptation in speech recognition~\cite{khurana2021unsupervised}.

\AtNextBibliography{\small}
\section{REFERENCES}
\label{sec:refs}
\printbibliography[heading=none]

\end{document}